# Adversarial Machine Learning-Based Anticipation of Threats Against Vehicle-to-Microgrid Services

Ahmed Omara, *Student Member, IEEE*, and Burak Kantarci, *Senior Member, IEEE*

*Abstract*—In this paper, we study the expanding attack surface of Adversarial Machine Learning (AML) and the potential attacks against Vehicle-to-Microgrid (V2M) services. We present an anticipatory study of a multi-stage gray-box attack that can achieve a comparable result to a white-box attack. Adversaries aim to deceive the targeted Machine Learning (ML) classifier at the network edge to misclassify the incoming energy requests from microgrids. With an inference attack, an adversary can collect real-time data from the communication between smart microgrids and a 5G gNodeB to train a surrogate (i.e., shadow) model of the targeted classifier at the edge. To anticipate the associated impact of an adversary's capability to collect real-time data instances, we study five different cases, each representing different amounts of real-time data instances collected by an adversary. Out of six ML models trained on the complete dataset, K-Nearest Neighbour (K-NN) is selected as the surrogate model, and through simulations, we demonstrate that the multi-stage gray-box attack is able to mislead the ML classifier and cause an Evasion Increase Rate (EIR) up to 73.2% using 40% less data than what a white-box attack needs to achieve a similar EIR.

*Index Terms*—vehicle-to-microgrid (V2M), machine learning, smart microgrids, cybersecurity, conditional generative adversarial network (CGAN), evasion attack, inference attack

## I. INTRODUCTION

Smart microgrids are empowered with communicational actuators and sensors that leverage the sustainability and resiliency factors of the power grid. Those factors call for proactive emergency preparedness and self-recovery solutions as proposed in Community Resilient Microgrids (CRM) [1]. The purpose of CRMs is to enhance the availability and sustainability of the delivered power, especially when the main grid is unavailable due to natural disasters and severe weather conditions. Hence, in order to sustain the CRM's goals amid power outages, the concept of energy trading, which we refer to as Vehicle-to-Microgrid (V2M, a.k.a V2MG), builds on utilizing the Electric Vehicles' (EVs) batteries as mobile energy units.

Smart microgrids rely on wireless communication devices to remotely connect with the existing power grid [2]. Wireless devices use data interfaces to communicate with other devices on a local power grid such as substations and transformers. These interfaces make them susceptible to being hacked by sophisticated technology trying to modify the behavior of the device or cause physical damage to it. Over the past few years, we have already seen the rise of malicious actors targeting smart grids and EVs' infrastructure [3].

The authors are with the School of Electrical Engineering and Computer Science at the University of Ottawa, Ottawa, ON, K1N 6N5, Canada. E-mail: {aomar020,burak.kantarci}@uottawa.ca

Although the cybersecurity threats under V2M settings have not been studied heavily, some of the existing research efforts investigated the cyber-attacks against the delivery and transmission of data in vehicular networks. For instance, the authors in [4] used an active control concept to detect a special type of data integrity attack, namely False Data Injection (FDI) attacks in a vehicular platoon. A data trust framework is proposed to verify the trustworthiness of the transmitted data among the vehicles [5]. Moreover, in our previous work, we studied the impacts of data integrity attacks against V2M applications [6].

This paper studies the impact of potential attacks against V2M services built upon adversarial machine learning. This attack consists of multiple stages where the adversary first launches an inference attack to train a surrogate model to perform an evasion attack. The inference attack uses a deep neural network that generates synthetic requests which are indistinguishable from real microgrid requests. The adversary uses a unique type of generative model, namely a Conditional Generative Adversarial Network (CGAN) [7], that is trained to learn to synthesize data samples that are statistically similar to real data samples. To the best of our knowledge, for the first time, this paper studies a gray-box evasion attacks in V2M settings. The main contributions are as follows:

- Present an anticipatory study of a multi-stage AML-based attack in V2M settings consisting of an inference attack and an evasion attack. We also provide possible directions on how to prevent such AML-based attacks.
- Demonstrate the associated impact of an adversary collecting different amount of real-time data instances on V2M services. We show that a white-box attack, where an adversary has access to 100% of the training dataset of the targeted classifier, an adversary can significantly damage V2M services with an Adversarial Detection Rate (ADR) of 2.4% and EIR of 97.5%. In addition, we show how in a gray-box attack, an adversary can use CGAN to generate synthetic data to augment the few collected real-time data instances. We anticipate significant damage at an ADR of 4.8% and EIR of 95.1%.

The rest of the paper is organized as follows. Section II provides an overview on adversarial machine learning attacks, particularly inference and evasion attacks. Section III explains the anticipated attacks and the implementation of CGAN. Section IV demonstrates performance criteria and numerical results regarding with the anticipated attacks. Section V concludes the paper and provides future directions.

## II. ADVERSARIAL MACHINE LEARNING

In this paper, we focus on two main types of adversarial machine learning attacks that occur during inference (test) time, namely inference attacks and evasion attacks. We study the impact of both attacks as they are potential venues of an adversary against V2M services. Hence, throughout this paper we seek answers to the following research question: what is the impact of AML-based attacks against V2M services if an adversary has limited knowledge of the V2M service? Thus, the motivation is to have better understanding of the vulnerabilities in a V2M system that are likely to be exploited by an adversary. Knowing these vulnerabilities will help defend V2M services against AML-based attacks by creating powerful mitigation systems as well as early-detection techniques.

### A. Inference (Exploratory) Attacks

The aim of inference attacks is creating a surrogate (i.e., shadow) model of the machine learning algorithm that resides at the targeted system. The surrogate model replicates the statistical distributions and functionalities of the targeted model. The adversary uses the surrogate model to infer the targeted model statistics. Consequently, the adversary can launch other attacks such as integrity attacks, that aim at misclassifying the inputs or reducing the targeted model's confidence.

Based on the availability of the targeted model to the adversary, the inference attack can be categorized into three classes, namely (1) white-box, (2) black-box and (3) gray-box attacks. In white-box attacks, the targeted model is known to the adversary and can be easily replicated using the models hyper-parameters and the training dataset. In black-box attacks, the adversary cannot access neither the models hyper-parameters nor the training dataset to create a surrogate model. In gray-box attacks, the adversary has access to some of the targeted model's hyper-parameters and the training dataset. The surrogate models performance depends on the quality and quantity of the captured observations. For instance, the adversary can query the targeted model and the corresponding labels to build a dataset [8], [9]. However, it might not be possible for the adversary to capture enough observations to train a high quality surrogate model.

### B. Evasion Attacks

Another type of attack that can occur during the testing time is evasion attacks. The adversary's goal is to feed the targeted model with adversarial examples (i.e., samples) to significantly reduce the model's integrity. The carefully perturbed samples are designed to deceive the model into making wrong decisions. The adversary chooses the designed samples based on the output labels' distances to the decision boundaries. Output labels adjacent to the decision boundaries tends to increase the likelihood of mis-classification at the targeted model. To design highly deceiving adversarial examples, the adversary has to study the targeted model by performing an inference attack.

## III. ANTICIPATION OF AML-BASED ATTACKS AGAINST V2M SERVICES

In this section we present an anticipatory framework to assess the impact of AML-based attacks that can be potentially launched against V2M services. In consideration of the involved components, the following two assumptions are made:

(i) An adversary in a V2M setting can directly obtain the output (i.e., classification label) of the targeted machine learning classifier at the mobile edge.

(ii) An adversary can manipulate the input data (i.e., observations/features) of the targeted machine learning classifier, and perform a data integrity attack as demonstrated in our previous work [6]. In addition, the authors in [10] showed that data integrity attacks are likely to occur by exploiting vulnerabilities of Advance Meter Infrastructure (AMI). The authors presented a threat model with four different falsification modes where an adversary can change and manipulate AMI's inputs.

Before discussing the potential impact of AML-based attacks, it is worth covering the interaction among the three main entities in a V2M setting, namely EVs, smart microgrids and gNodeB. Smart microgrids, that are predicted to experience a power outage, send service requests to the gNodeB. The machine learning model at the gNodeB classifies the incoming requests into high-priority, medium-priority or low-priority requests. The three priority groups represent the level of energy necessity for microgrids, taking into consideration power generation and power consumption factors. The requests are labeled based on the input features sent by the smart microgrids [11]. The gNodeB broadcasts the request to the EVs within its coverage range. The EVs respond to the request with the following: (i) selling price, (ii) contribution percentage of the EVs battery, and (iii) current location. Then, the gNodeB calculates distances between the microgrids and the EVs, and chooses an optimal set of EVs to serve each microgrid's request. Lastly, the gNodeB sends the trajectories to the EVs [12].

In the following subsections, we discuss AML-based attacks, and their use of CGAN to generate synthetic data in the context of V2M settings.

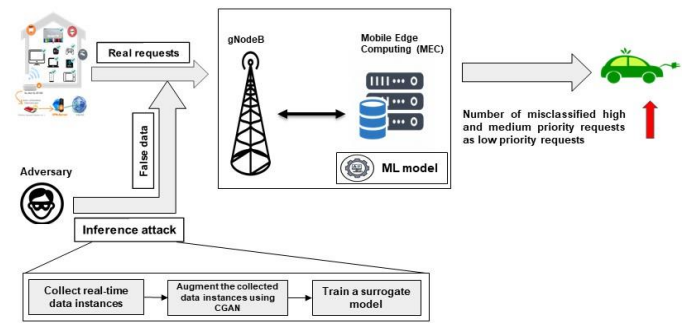

Fig. 1. An overview of anticipated attacks on V2M services

## A. Inference and Evasion Attacks Against V2M Services

The anticipated inference attack launched against the discussed V2M framework consists of three steps as depicted in Fig. 1. First, an adversary collects sufficient real-time data instances (i.e., observations) and output data (i.e., priority labels). It is shown that microgrids' communication systems are susceptible to cybersecurity attacks which can lead to data leaks [13]. Collecting enough observations is a difficult task for the adversary. Hence, the second step is using the CGAN model to generate synthetic data which, with addition to the observations, builds a complete dataset. Third, the adversary trains different surrogate models on the complete dataset, and the model with the best performance, specifically accuracy and F1-score, is selected as the optimal surrogate model. With the help of the surrogate model, the adversary launches an evasion attack where they inject adversarial samples into the microgrid's real requests. As a result, the operating ML classifier at the mobile edge classifies the high and medium priority requests as low priority requests.

## B. Conditional Generative Adversarial Network

To augment the collected observations (i.e., real-time data instances), the adversary uses a CGAN to generate synthetic data. However, with too few collected observations, the CGAN cannot accurately model the distribution of the real data. Hence, we investigate the impact of different amounts of collected observations on the CGAN performance. This paves the way for unveiling the potential vulnerabilities of a protected V2M system that can still be exploited, and leads to solid guidelines for the design of resilient techniques to mitigate the anticipated damage with possible prevention of adversarial attacks.

Generative Adversarial Networks (GANs) can be used to produce synthetic data that resembles the real data input to the networks. CGANs can use data labels during the training process to generate data belonging to specific categories. Additional information that is correlated with the input data, such as class labels, can be used to improve GAN performance. For instance, synthetic data can be generated in a better quality while maintaining more stable or faster training. CGANs are trained in such a way that both the generator and the discriminator models are conditioned on the class label, so that when the trained generator is used as a standalone model to generate samples in the domain, samples of a given type can be generated. CGANs consist of two networks that train together as adversaries:

Generator network - Given a label and random array as input, this network generates data with the same structure as the training data observations corresponding to the same label. The objective of the generator is to generate synthetic-labeled data that the discriminator classifies as "real."

Discriminator network - Given batches of labeled data containing observations from both training data and generated data from the generator. The discriminator tries to classify the instances as "real" or "synthetic". The objective of the discriminator is to not be deceived by the generator when given batches of both real and synthetic (i.e., generated) labeled data. The entire process corresponds to a minimax game played between $D$ and $G$ as stated in equation (1) [14]:

$$\min_G \max_D \mathbb{E}_{x,y \sim p_{data}(x,y)} log[D(x,y)] + \\ \mathbb{E}_{y \sim p_Y(y), z \sim p_Z(z)} log[1 - D(G(z,y),y)] \quad (1)$$

, where $y$ represents the extra information (i.e., labels) drawn from a distribution $p_Y(y)$.

## IV. PERFORMANCE EVALUATION

### A. Experiment Setup

To assess the impact of inference and evasion attacks on V2M services, MATLAB Deep Learning and Simscape Electrical toolboxes are used. All simulations are performed using Intel Core i7-10700F CPU with 32GB of RAM and NVIDIA GeForce GTX 1660 SUPER GPU running on a Windows 10 system. The Deep Learning Toolbox is used to run the CGAN model; whereas, the Simscape Electrical Toolbox simulates the EVs [15].

*1) Dataset preparation and mobile edge-based classifier:* We use the iHomeLab RAPT dataset to run our experiments [16]. The dataset consists of electrical power consumption data as well as power generation data for five residential households in Switzerland. The power consumption data can be found in two resolutions: appliance-tier data and aggregated household-tier data. The power generation data comes from Photovoltaic panels (PVs). The data was collected in a period of 1.5 to 3.5 years with different sampling frequency for each household. In this paper, we chose one residential household from the iHomeLab RAPT dataset with a sampling frequency of 10 minutes. In addition, we supplement the iHomeLab RAPT dataset with a previously used dataset in [6]. The supplement dataset comprises of a power generation module. The power generation module, which uses a wind farm and back-up batteries as alternative resources of power, contains power generation data such as wind farm's generation profile and the capacity of the back-up batteries.

We use K-means, an unsupervised machine learning algorithm, to cluster the aggregated dataset (D), which consists of iHomeLab RAPT and our own dataset, into three priority groups, namely low-priority, medium-priority and high-priority. Each one of the three priority groups represents the level of energy requirement for the microgrids; the priority groups take energy generation and power consumption factors into consideration. Hence, K-means clusters the dataset based on the power consumption and the energy generation profiles as in [17].

After clustering and labeling the aggregated dataset, we use the dataset (D) to train a K-NN model to operate as an ML classifier at the network edge. K-NN was selected as it outperforms other ML classifiers that were trained on the labeled dataset such as Support Vector Machine (SVM) and Logistic Regression (LR).

*2) Inference attack's settings:* As previously mentioned, the inference attack starts with an adversary collecting real-time data instances to build a training dataset. To provide a comprehensive analysis, we compare five cases, where an adversary has access to different amounts of data observations. For the white-box attack, the adversary has full access to the labeled (training) dataset (D) of the operating ML classifier at the edge. For the gray-box attack, the adversary can collect limited real-time data instances from the communication between the microgrids and gNodeB, namely 1,600, 1,200, 800 and 400 which corresponds to 80%, 60%, 40% and 20% of the training dataset's size.

In the gray-box attack, after obtaining the data instances (i.e., data observations), the adversary complements the few collected observations using CGAN to generate a complete dataset. To perform a fair comparison, we assume that CGAN generates enough data to complement the missing percentage of the dataset, such that the adversary will always have the same dataset's size in all cases. For instance, if the adversary collects 1,800 data instances (i.e., 80% of the training dataset's size), CGAN is used to generate the remaining 200 data instances (i.e., 20% dataset's size) as illustrated in Table I. For the white-box attack, where the adversary has 100% access to the training set, the data augmentation step is skipped. The CGAN's architecture is depicted in Fig. 2 where the generator and the discriminator are explained in detail. A brief description of the generator operations can be summarized in the following four points: projecting and reshaping the random noise, converting the categorical labels to embedding vectors, concatenating the outputs of the random noise and labels, and upsampling the concatenated arrays using a series of transposed convolution, batch normalization and Rectified Linear Unit (ReLU) layers. The discriminator operates similarly except for the last point, where the discriminator downsamples the concatenated arrays using a series of convolution layers with leaky ReLU layers.

The last part of the inference attack is to choose a surrogate model that captures the underlying functionalities of the targeted classifier at the mobile network edge. Six different models are trained on the complete dataset (i.e., real and synthetic data). The six models that we consider are Decision Tree (DT), Random Forest (RF), Logistic Regression (LR), K-NN, SVM and Naive Bayes (NB). The models are trained and tested on the 80% and 20% partition of the complete dataset, respectively. The model with the best accuracy and F1-score performances is selected as the surrogate model. It is worth to mention that the test set (i.e., 20%) consisted of the real-time data instances collected by the adversary, not data generated by the CGAN. Hence, this allows us to assess the capability of a CGAN model to capture the distribution of the real-time data.

*B. Evaluation Metrics*

In this section, we introduce different metrics to assess the potential damage of the inference attack and the evasion attack on V2M services. The impact of the inference attack is assessed based on the surrogate's model performance, particularly on the testing data. High performance of the surrogate model demonstrates the effectiveness of the CGAN's model to generate high data quality that mimics the statistical distribution of the real-time data. We evaluate the six models in terms of accuracy ($(TP + TN)/(TP + FP + TN + FN)$) and F1-score ($2 \cdot (Precision \cdot Recall)/(Precision + Recall)$) where $Precision = \frac{TP}{TP+FP}$ and $Recall = \frac{TP}{TP+FN}$. TP, TN, FP and FN stands for True Positive, True Negative, False Positive and False Negative, respectively.

The anticipated impact of the evasion attack is determined based on the adversary's ability to deceive the K-NN classifier at the gNodeB such as the number of high-priority and medium-priority requests that are falsely classified as low-priority requests. Two metrics are considered to evaluate the evasion attack: Evasion Increase Rate (EIR) and Adversarial Detection Rate (ADR) (i.e., True Negative Rate (TNR)).

$$ADR = \frac{TN}{TN + FP}$$

$$EIR = 1 - \frac{TNR_{adversarial}}{TNR_{original}}$$

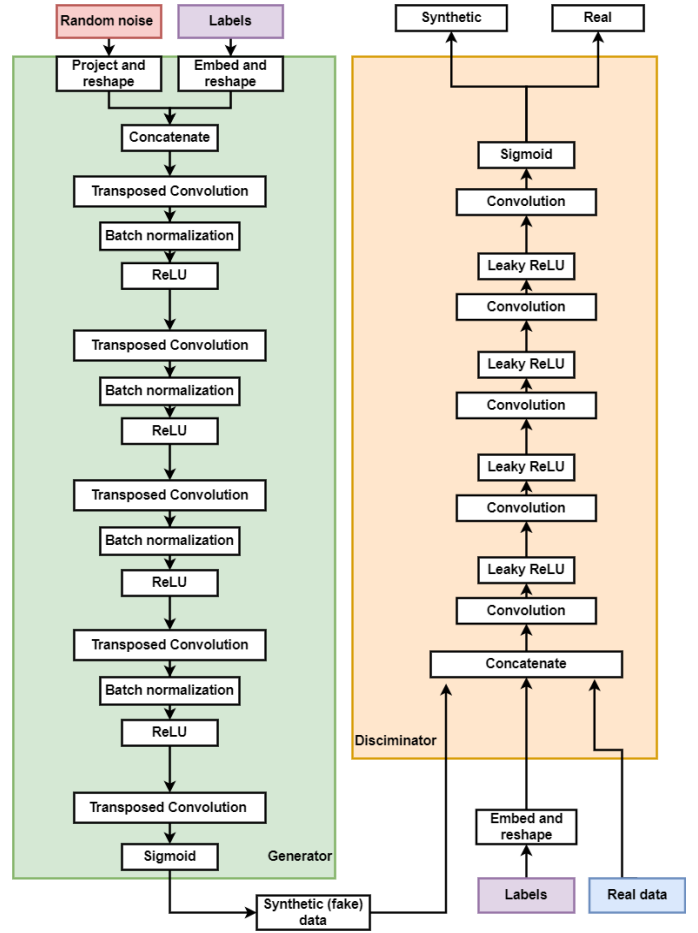

Fig. 2. Generator and discriminator architectures of CGAN

The $TNR_{adversarial}$ defines the adversarial detection rate, whereas $TNR_{original}$ defines the original detection rate of the classifier before the attack.

*C. Numerical Results*

We first start with the numerical results of the inference attack. Case-A is the white-box attack, where the adversary has access to 100% of the training dataset of the K-NN classifier at the mobile network edge. Cases B, C, D and E correspond to 80%, 60%, 40% and 20% of real-time data instances, respectively.

As the adversary has access to the complete training dataset in case-A, the synthetic data generation using CGAN is skipped as shown in Table I. In all other cases, the CGAN is built using the available data instances collected by the adversary. After the CGAN converges, the CGANs generator is used to generate data, which is used to complement the missing percentage of the collected dataset, creating a final combined dataset, consisting of collected and generated data. We ensure that all cases have a combined dataset of the same size. The combined dataset is used to train and test the surrogate models. To guarantee fair comparison, all surrogate models are trained and tested under 80% and 20% of the combined dataset, respectively. In addition, we use a balanced dataset, and the results of the surrogate models represent the average of 10 different runs.

TABLE I
DESCRIPTION OF THE COMBINED DATASET

| Case | | Combined dataset | |
|---|---|---|---|
| | | Adversary's contribution to the dataset (real-time collected data) | CGAN's contribution to the dataset (generated data) |
| White-box | A | 100% | 0% |
| Gray-box | B | 80% | 20% |
| | C | 60% | 40% |
| | D | 40% | 60% |
| | E | 20% | 80% |

Table II presents the anticipated performance of the surrogate model as a part of an inference attack against V2M services. In case-A, case-B and case-D, the adversary is able to choose a surrogate model similar to the target edge classifier (i.e., K-NN). For case-A, based on 98.35% accuracy and 97.5% F1-score, K-NN is selected as the surrogate model. Similarly, in case-B, the adversary was again able to select the same operating classifier at the edge. Even with decreasing the available real-time data instances from 100% to 80%, K-NN can still achieve 95.46% accuracy and 94.9% F1-score.

In case-C, with 60% of available data, SVM outperforms the other models and serves as the surrogate model. Despite SVM achieving an accuracy of 89.88% and an F1-score of 90%, K-NN performed slightly lower than SVM with an accuracy of 89.65% and F1-score 89.5%. In case-D, although the adversary only collects 40% of the real-time data instances, the performance in terms of accuracy and F1-score are acceptable, yielding 81.4% and 82.3%, respectively. The K-NN is selected as the surrogate model which matches the targeted classifier. In case-E, with 20% collectable data instances, the performance drops significantly leading to 70.15% accuracy and 69% F1-score under LR as the surrogate model. It is noted that in case-E, the number of collected real-time data instances are insufficient for the CGAN model to capture the underlying statistics of the data. Thus, when the LR classifier is tested on the real-time data instances, it demonstrates low performance as compared to the other cases. In addition, K-NN is not selected as the surrogate model because of the insufficient data instances.

TABLE II
ANTICIPATED ACCURACY AND F1-SCORE OF THE SURROGATE MODELS AS PART OF INFERENCE ATTACK (WHEN USING CGAN IN CASES B-E)

| Case | | Selected surrogate model | Performance | |
|---|---|---|---|---|
| | | | Accuracy | F1-score |
| White-box | A | K-NN | 98.35% | 97.5% |
| Gray-box | B | K-NN | 95.46% | 94.9% |
| | C | SVM | 89.88% | 89.43% |
| | D | K-NN | 81.4% | 80.9% |
| | E | LR | 70.15% | 69.2% |

It is observed that as the adversary has access to a smaller partition of the real-time dataset, the performance of the surrogate model decreases. It is worth noting that the accuracy and F1-score drop is logarithmic, not linear. For instance, between case-A and case-B, the accuracy decreases by 3% and the F1-score decreases by 4%; whereas, between case-D and case-E, the accuracy decreases by 11% and the F1-score decreases by 12%.

Results of the evasion attack are presented in Table III. Performance of case-A is significantly high with an ADR of 2.4% and an EIR of 97.5%. The reason behind that high impact of damage is because of the adversary's complete access (i.e., white-box) to the training dataset of the K-NN classifier at the edge. The performance starts dropping for the gray-box attacks as the adversary collects limited real-time data instances. For instance, in case-B, although only 80% data instances were collected, we notice that the adversary is able to deceive the gNode's K-NN classifier with ADR of 4.8% and EIR of 95.1%. However, the anticipated damage of the attack decreases significantly for case-C with an ADR of 20.6% and an EIR of 79.3%. This significant decrease is due to the mis-matching between the selected surrogate model (i.e., SVM) and the operating classifier at the network edge (i.e., K-NN). In addition, the boundary regions of the SVM for the low-priority requests are different than the K-NN, which might misguide the adversary about the feature values that lead to incremental false positives. On the contrary, in case-D, as the selected surrogate model matches the operating classifier, we observe a less significant decrease in the ADR and EIR results with 26.5% and 73.2%, respectively. Thus, the adversary is able to perform an impactful evasion attack by collecting only 40% of the real-time data instances and generating 60% synthetic data using CGAN. However, a significant decrease occurs under case-E due to two reasons. First, with only 20% of collected real-time data instances, the CGAN is unable to generate high quality synthetic data as it cannot capture the statistical distribution of the real-time data. Second, since the decision boundaries of the selected surrogate model (i.e., LR) and the operating K-NN classifier are different, the adversary cannot design adversarial perturbed samples that can successfully deceive the classifier. In Table IV, we show the potential impact of the evasion attack without using CGAN. Thus, without using CGAN, the impact of the ADR and EIR are reduced by at least 7% when compared to not using CGAN. In addition, we notice that the EIR of case-D with CGAN (73.2%) is higher than the EIR of case-C without CGAN (69.9%). In other words, augmenting 40% of real-time data instances (case-D) with CGAN results in an EIR of 73.2%, whereas 60% of real-time data instances (case-C) results in an EIR of 69.9% when not using CGAN. Hence, an adversary can potentially pose more damage to V2M services when collecting fewer real-time data instances and augmenting them with CGAN.

TABLE III
ANTICIPATED ADR AND EIR OF EVASION ATTACK AGAINST V2M SERVICES (WHEN USING CGAN IN CASES B-E)

| Case | | Selected surrogate model | Performance | |
|---|---|---|---|---|
| | | | ADR | EIR |
| White-box | A | K-NN | 3.4% | 96.5% |
| Gray-box | B | K-NN | 4.8% | 95.1% |
| | C | SVM | 20.6% | 79.3% |
| | D | K-NN | 26.5% | 73.2% |
| | E | LR | 43.5% | 56.4% |

TABLE IV
ANTICIPATED ADR AND EIR OF EVASION ATTACK AGAINST V2M SERVICES (WITHOUT USING CGAN)

| Gray-box Case | Performance | |
|---|---|---|
| | ADR | EIR |
| B | 11.7% | 88.1% |
| C | 29.9% | 69.9% |
| D | 40.7% | 59.2% |
| E | 59.3% | 40.5% |

## V. CONCLUSION

In this paper, we presented an anticipatory study of a multi-stage adversarial machine learning attack in Vehicle-to-Microgrid (V2M) settings, consisting of an inference attack and an evasion attack. Knowledge of the adversary has been taken into consideration as we quantify the impact of the collectable real-time data instances. We have compared five different cases of combined dataset, one case under white-box setup and four cases under gray-box setup. In the white-box case, the adversary has 100% access to the training dataset of the K-NN classifier; whereas, in the gray-box cases, we have assumed that the adversary collects different amounts of real-time data instances, and the rest of the dataset is complemented using a CGAN model. Through simulations, we have presented the selected surrogate model for each case based on the accuracy and F1-score performance. As the number of collected data decreases, the surrogate model's performance drops logarithmically. In addition, we have shown that an adversary is able to deceive the K-NN classifier at the edge, achieving high ADR and EIR when the selected surrogate model is K-NN, matching the operating classifier. Moreover, we have anticipated that with 40% collected data, the potential damage of evasion attack results in 26.5% ADR and 73.2% EIR. We have concluded that, with the lack of powerful mitigation techniques, AML-based attacks on V2M services can significantly damage the system.

ACKNOWLEDGMENT

This work was supported in part by the Natural Sciences and Engineering Research Council of Canada (NSERC) under the DISCOVERY Program.